\documentclass[aps,prl,reprint,groupedaddress,superscriptaddress]{revtex4-1}
\usepackage{graphicx,color,bm,amsmath,amssymb,url,natbib} 

\newcommand{\Tc}{T_{\text c}}
\newcommand{\Cs}{CsV$_3$Sb$_5$}
\newcommand{\CsTax}{Cs(V$_{1-x}$Ta$_x$)$_3$Sb$_5$}
\newcommand{\CsTa}{Cs(V$_{0.86}$Ta$_{0.14}$)$_3$Sb$_5$}

\begin{document}

\title{Conventional Superconductivity in the Doped Kagome Superconductor Cs(V$_{0.86}$Ta$_{0.14}$)$_3$Sb$_5$ from Vortex Lattice Studies}

\author{Yaofeng~Xie}
\altaffiliation{These authors contributed equally to the work}
\affiliation{Department of Physics and Astronomy, Rice University, Houston, Texas 77005, USA}

\author{Nathan~Chalus}
\altaffiliation{These authors contributed equally to the work}
\affiliation{Department of Physics and Astronomy, University of Notre Dame, Notre Dame, Indiana 46556, USA}

\author{Zhiwei~Wang}
\altaffiliation{These authors contributed equally to the work}
\affiliation{Centre for Quantum Physics, Key Laboratory of Advanced Optoelectronic Quantum Architecture and Measurement (MOE), School of Physics, Beijing Institute of Technology, Beijing 100081, China}
\affiliation{Beijing Key Lab of Nanophotonics and Ultrafine Optoelectronic Systems, Beijing Institute of Technology, Beijing 100081, China}
\affiliation{Material Science Center, Yangtze Delta Region Academy of Beijing Institute of Technology, Jiaxing, 314011, China}

\author{Weiliang~Yao}
\affiliation{Department of Physics and Astronomy, Rice University, Houston, Texas 77005, USA}

\author{Jinjin~Liu}
\affiliation{Centre for Quantum Physics, Key Laboratory of Advanced Optoelectronic Quantum Architecture and Measurement (MOE), School of Physics, Beijing Institute of Technology, Beijing 100081, China}
\affiliation{Beijing Key Lab of Nanophotonics and Ultrafine Optoelectronic Systems, Beijing Institute of Technology, Beijing 100081, China}

\author{Yugui~Yao}
\affiliation{Centre for Quantum Physics, Key Laboratory of Advanced Optoelectronic Quantum Architecture and Measurement (MOE), School of Physics, Beijing Institute of Technology, Beijing 100081, China}
\affiliation{Beijing Key Lab of Nanophotonics and Ultrafine Optoelectronic Systems, Beijing Institute of Technology, Beijing 100081, China}
\affiliation{Material Science Center, Yangtze Delta Region Academy of Beijing Institute of Technology, Jiaxing, 314011, China}

\author{Jonathan~S.~White}
\affiliation{Laboratory for Neutron Scattering and Imaging (LNS), PSI Center for Neutron and Muon Sciences, Paul Scherrer Institute, 5232 Villigen PSI, Switzerland}

\author{Lisa~M.~DeBeer-Schmitt}
\affiliation{Large Scale Structures Section, Neutron Scattering Division, Oak Ridge National Laboratory, Oak Ridge, Tennessee 37831, USA}

\author{Jia-Xin Yin}
\affiliation{Department of Physics, Southern University of Science and Technology, Shenzhen 518055, People’s Republic of China}

\author{Pengcheng~Dai}
\affiliation{Department of Physics and Astronomy, Rice University, Houston, Texas 77005, USA}

\author{Morten Ring~Eskildsen}
\affiliation{Department of Physics and Astronomy, University of Notre Dame, Notre Dame, Indiana 46556, USA}

\begin{abstract}
A hallmark of unconventional superconductors is a complex electronic phase diagrams where intertwined orders of charge-spin-lattice degrees of freedom compete and coexist.
While the kagome metals such as CsV$_3$Sb$_5$ also exhibits complex behavior, involving coexisting charge density wave order and superconductivity, much is unclear about the microscopic origin of the superconducting pairing.
We study the vortex lattice in the superconducting state of Cs(V$_{0.86}$Ta$_{0.14}$)$_3$Sb$_5$, where the Ta-doping suppresses charge order and enhances superconductivity.
Using small-angle neutron scattering, a strictly bulk probe, we show that the vortex lattice exhibits a strikingly conventional behavior.
This includes a triangular symmetry with a period consistent with 2$e$-pairing, a field dependent scattering intensity that follows a London model, and a temperature dependence consistent with a uniform superconducting gap.
Our results suggest that optimal bulk superconductivity in Cs(V$_{1-x}$Ta$_x$)$_3$Sb$_5$ arises from a conventional Bardeen-Cooper-Schrieffer electron-lattice coupling, different from spin fluctuation mediated unconventional copper- and iron-based superconductors.
\end{abstract}

\date{\today}

\maketitle

\section{Introduction}
n conventional Bardeen-Cooper-Schrieffer (BCS) superconductors, the electron-lattice coupling leads to the formation of coherent ($2e$) Cooper pairs and the opening of an isotropic $s$-wave gap at the Fermi level \cite{Bardeen:1957hw}.
In comparison, a key signature of unconventional superconductivity in materials such as copper oxides and iron-pnictides is that the pairing may be mediated by spin fluctuations \cite{Scalapino:2012ek,Dai:2015wj} and associated with  intertwined charge-spin-lattice degrees of freedom~\cite{Fradkin:2015wv}.
The discovery of superconductivity in the layered kagome $A$V$_3$Sb$_5$ ($A$ = K, Rb, Cs) metals
\cite{Ortiz:2020by,Jiang.2022,Wilson.2024z} is interesting because the superconducting state develops in the presence of a charge density wave (CDW)~\cite{Jiang:2021tw,Liang:2021ft,Li:2021we,Zhao.2021,Chen:2021vu,Kang:2022vk}, and the competition between these two ordered states may give rise to unconventional superconductivity~\cite{Tan:2021uj,Xu:2021vc,Song:2021ww,Chen.2022,Kang.2023}.
However, no spin fluctuations are reported, thus raising the question whether these materials are conventional BCS or unconventional superconductors like cuprates and iron pnictides.
From pressure and Ta-doping dependence of the {\CsTax} phase diagram, it is clear that CDW order competes with superconductivity, and optimal superconductivity of $T_c = 5.3$ K appears around {\CsTa} with vanishing CDW order~\cite{Zhong.2023,Xiao.202349}.

The majority of experimental evidence in the $A$V$_3$Sb$_5$ materials indicate singlet $s$-wave pairing with a nodeless but possibly anisotropic superconducting gap~\cite{Mu:2021dl,Duan:2021bf,Roppongi.2023,Zhong.2023}.
Furthermore, multiband superconductivity with a large difference in the gap size on different Fermi surface sheets have been reported~\cite{Duan:2021bf,Chen:2021vu,Xu:2021vc,Yin.2021,Gupta:2022vb,Roppongi.2023}.
An experimental determination of electron-lattice coupling strength supports conventional BCS superconductivity~\cite{Zhong.20239gn}, although there are questions whether the relation between the gap and critical temperature is consistent with weak coupling~\cite{Tan:2021uj}.
There are several reports of broken time-reversal symmetry (BTRS) both in the normal state~\cite{Jiang:2021tw,Feng:2021wr,Khasanov.2022,Hu.2022lu} as well as the superconducting state~\cite{MielkeIII:2022wd,Guguchia.2023,Le.2024}, although it is not observed consistently in all experiments~\cite{Roppongi.2023,Saykin.2023,Wang.2024srn}.
In addition, possible Majorana bound states were observed in scanning tunneling spectroscopy studies~\cite{Liang:2021ft}.
Finally, a transition from conventional $2e$-pairing towards (vestigal) $4e$- and $6e$-pairing~\cite{Herland.2010} has been reported upon heating towards the transition to the normal state~\cite{Zhou.2022k9,Hecker.2023,Zhang.20223eb,Ge.2024,Varma.2023}.

\begin{figure*}
    \includegraphics{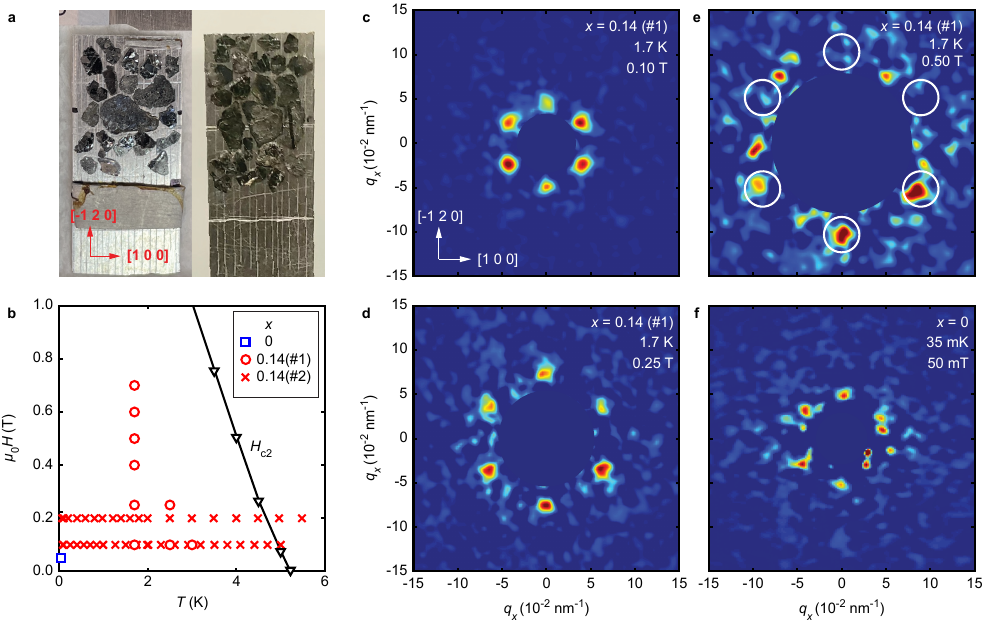}
    \caption{\label{SampleDifPats}
        {\bf Vortex lattice diffraction patterns.}
        {\bf a} Cs(V$_{0.86}$Ta$_{0.14}$)$_3$Sb$_5$ sample mosaic used for the SANS experiments, with the arrows indicating the in-plane crystalline directions.
        {\bf b} Field-temperature phase diagram indicating where SANS were performed.
        {\bf c} Cs(V$_{0.86}$Ta$_{0.14}$)$_3$Sb$_5$ VL diffraction pattern at 1.7~K and 0.1~T.
        The arrows indicate the crystalline orientation.
        {\bf d} Same at 0.25~T and {\bf e} 0.5~T.
        In panel (e) only Bragg peaks at the lower half of the detector were rocked through the Ewald sphere and the positions of symmetry equivalent peaks are indicated by circles.
        {\bf f} CsV$_3$Sb$_5$ VL diffraction pattern at 35~mK and 50~mT.
        For all diffraction patterns, background measurements obtained at zero field are subtracted and the region near $q = 0$ is masked off.
        A separate color scale is used for each diffraction pattern.
        }
\end{figure*}

To determine the microscopic origin of superconductivity in $A$V$_3$Sb$_5$, it is critical to separate effects of the CDW as this may be associated with an electronic nematic phase and affect the electron pairing \cite{Nie.202222,Asaba.2024}.
Here, we report on small-angle neutron scattering (SANS) studies of the vortex lattice (VL) induced by an applied magnetic field in {\CsTa} where charge ordering is suppressed~\cite{Zhong.2023,Xiao.202349}.
The vortices produce singularities in the order parameter, and may be used as probes of the superconducting state in the host material~\cite{Muhlbauer2019,White.2021}.
Our results indicate a highly conventional bulk superconducting state, and thus suggest that the reported exotic behavior in other members of the $A$V$_3$Sb$_5$ family of superconductors are not relevant to the microscopic origin of superconductivity.
Importantly, the SANS technique provides information about the bulk superconducting state, whereas more exotic phenomena such as BTRS may occur only at the sample surface.

\section*{Results}
\subsection*{Vortex lattice imaging}
The sample used for the SANS experiments, shown in Fig.~\ref{SampleDifPats}(a), consisted of a mosaic of co-aligned {\CsTa} single crystals. 
Two SANS experiments (\#1/2) were performed, exploring different temperature and field ranges as indicated in the phase diagram in Fig.~\ref{SampleDifPats}(b).
In all cases, the magnetic field was applied perpendicular to the six-fold symmetric (kagome) lattice planes.
Figures~\ref{SampleDifPats}(c)-\ref{SampleDifPats}(e) show VL diffraction patterns for {\CsTa} at three different applied magnetic fields.
The system was prepared by a field cooling to the measurement temperature from above $T_c$, followed by a damped field oscillation with an initial amplitude of 5\% of the measurement field.
As witnessed by the well-defined defined Bragg peaks, this produces an ordered VL which is consistent with weak pinning~\cite{Levett:2002ba}.

A triangular VL is observed at all fields and temperatures as expected for a superconductor with a six-fold symmetric basal plane, and oriented with Bragg peaks along the $[\overline{1} \, 2 \, 0]$ crystalline direction.
The same symmetry and orientation is observed in a reference measurement on undoped {\Cs} as shown in Fig.~\ref{SampleDifPats}(f), although VL imaging is only possible at a low field and temperature due to the much larger penetration depth ($182.7$~nm vs $106.8$~nm) and much lower upper critical field ($0.3$~T vs $2.2$~T)~\cite{Li.2022cod}.
In contrast, vortex imaging by scanning tunneling spectroscopy (STS) founds a VL that undergoes a $15^{\circ}$ rotation as the applied field is increased from 75~mT to 200~mT, and where the high field orientation is orthogonal to the one reported here~\cite{Huang.2024}.
This highlights how different results, reflecting the bulk properties, may be obtained using SANS compared to surface probes with a limited field of view such as STS.

The VL orientation is determined by anisotropies within the screening current plane, which may arise from the Fermi surface (FS)~\cite{Kogan:1997vm} or the superconducting gap~\cite{Franz:1997jn,Agterberg:1998wo,Ichioka:1999aa}.
Considering reports of an isotropic gap on all three FS sheets in {\CsTa}~\cite{Zhong.2023}, the former scenario is the most likely. 
A direct correlation between the VL orientation and the band structure requires an evaluation of Fermi velocity averages as well as a directionally resolved Density of States at the Fermi level~\cite{Kogan:1997vm,Hirano:2013jx}, which has presently not been carried out for {\Cs} or {\CsTa}.
Nevertheless, the morphology of almost perfectly nested hexagonal ($\beta$) and triangular ($\delta$) FS sheets with flat sections perpendicular to the $\Gamma$-$K$ direction~\cite{Ortiz:2021vb,Kang:2022vk,Hu:2022wz,Zhong.2023}, is consistent with the observed VL orientation.
This commonality, together with the isotropic ($\alpha$) FS sheet,  may also explain the absence of a field driven VL rotation transition observed in multiband superconductors such as MgB$_2$~\cite{Cubitt:2003ab} and UPt$_3$~\cite{Avers:2020wx,Avers.2022}.

\subsection*{Scattering vector magnitude}
\begin{figure}
    \includegraphics{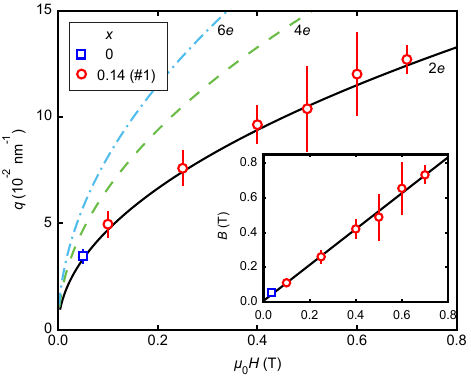}
    \caption{\label{FluxQuant}
         {\bf Field dependence of the vortex lattice scattering vector.}
         Full, dashed and dot-dashed lines indicate $q$ expected for $2e$-, $4e$- and $6e$-pairing respectively.
         Inset shows the magnetic induction determined from eqn.~(\ref{qne}) assuming $2e$-pairing.
         The line is a linear fit.
         Data for $x = 0.14$ was measured at $1.7$~K and for $x = 0$ at 35~mK.
         For both the main figure and the inset, error bars represent one standard deviation.
         }
\end{figure}

The scattering vector magnitude for a triangular VL is
\begin{equation}
  q_{ne}(B) = 2\pi \sqrt{\frac{2B}{\sqrt{3} \Phi_{ne}}},
  \label{qne}
\end{equation}
where $B$ is the magnetic induction, $\Phi_{ne} = h/ne$ is the flux quantum and $n$ is an even integer.
For regular $2e$-pairing the flux quantum is given by $\Phi_{2e} = \Phi_0 = 2068$~Tnm$^2$~\cite{Abrikosov:1957vu}.
Figure~\ref{FluxQuant} shows the measured VL scattering vector ($q$) versus applied magnetic field ($\mu_0 H$).
This follow the behavior expected for $2e$-pairing, assuming $B = \mu_0 H$.
Furthermore, $q$ can be reliably determined at temperature up to $\sim \tfrac{2}{3} T_c$ at $0.1$~T 
and shows no deviation from $q_{2e}$.
In contrast, $q_{4e}$ and $q_{6e}$ are not compatible with the data within the experimental error.
While agreement could in principle be achieved for $n \neq 2$, it requires $B = \sqrt{n/2} \, \mu_0 H$ which is inconsistent with magnetization measurements on {\CsTa}~\cite{Li.2022cod}.
The SANS results does thus not provide evidence for $4e$- and $6e$-pairing at the bulk level that has been reported for {\Cs} as one approaches $T_c$~\cite{Ge.2024}.
The inset to Fig.~\ref{FluxQuant} shows the magnetic induction inferred from the measured $q$ and using eqn.~(\ref{qne}) with $n = 2$.
A linear fit to the data yields a slope of $1.033 \pm 0.015$ and an ordinate intercept of $B = 7.4 \mbox{ mT} \pm 4.2 \mbox{ mT}$.
The latter provides an upper limit of a few milliTesla on any net spontaneous field.

\subsection*{Form factor field dependence}
\begin{figure}
    \includegraphics{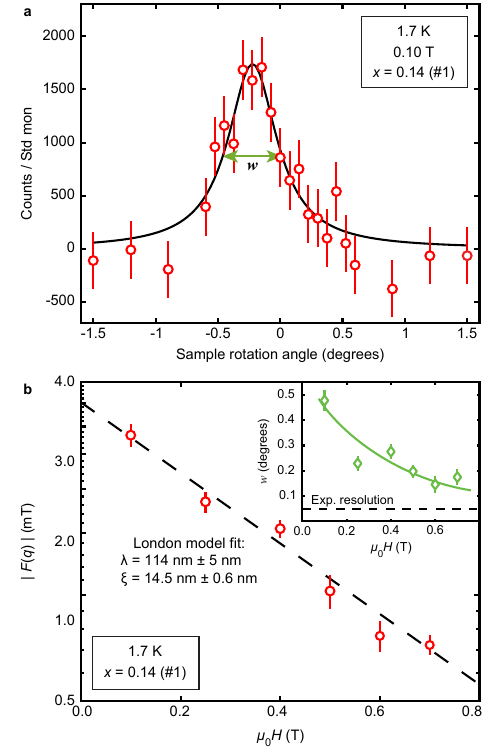}
    \caption{\label{FormFactor}
        {\bf Vortex lattice form factor.}
        {\bf a} Scattered intensity as the VL Bragg is rotated through the Ewald sphere.
        The curve is a Lorentzian fit and the full-width-half-maximum ($w$) is indicated by the arrow.
        {\bf b} VL form factor versus applied magnetic field.
        The dashed line is a fit to eqn.~(\ref{LondonFF}), with the penetration depth and coherence length indicated in the plot.
        Inset shows the full-width-half-maximum rocking curve widths obtained from Lorentzian fits.
        The solid line is a guide to the eye and the dashed line shows the experimental resolution.
        For all panels, error bars indicate one standard deviation.
        }
\end{figure}

The field dependence of the scattered intensity provides information about the superconducting penetration depth ($\lambda$) and coherence length ($\xi$).
This requires a measurement of the integrated intensity, obtained by rotating the VL diffraction peak through the Bragg condition as shown in Fig.~\ref{FormFactor}(a).
Normalizing by the incident neutron flux one obtains the VL reflectivity
\begin{equation}
  R = \frac{2 \pi \gamma_n^2 \, \lambda_n^2 \, t_s}{16 \Phi_{ne}^2 \, q} \left| F(q) \right|^2,
  \label{Refl}
\end{equation}
where $F(q)$ is the VL form factor, $\gamma_n = 1.913$ is the neutron magnetic moment in units of the nuclear magneton, and $t_s$ is the sample thickness~\cite{Kemoklidze:1965wi,Christen:1977aa}.
Figure~\ref{FormFactor}(b) shows the form factor obtained from the SANS measurement which is found to decrease exponentially with increasing field.
This is consistent with the London model
\begin{equation}
  F(q) = \frac{B}{1 + q^2 \lambda^2} \, e^{- q^2 \xi^2/2},
  \label{LondonFF}
\end{equation}
with a Gaussian core cut-off to account for a finite coherence length~\cite{Yaouanc:1997aa,Eskildsen:2011iv}.
The numerical factor of $\tfrac{1}{2}$ in the exponent has been found to yield reasonable values for the coherence length in a range of superconductors, although other values has been used in the literature~\cite{Muhlbauer2019}.
For $q\lambda \gg 1$ the only field dependence is through $q^2 \propto B$ in the exponent.
A fit to the data yields $\lambda = (113.7 \pm 4.8)~\mbox{nm}$ from the zero field intercept, and $\xi = (14.5 \pm 0.6)~\mbox{nm}$ from the slope.
This agrees well with values for $\lambda = 106.8$~nm and $\xi = 12.2$~nm inferred from measurements of the lower and upper critical fields~\cite{Li.2022cod}.
Notably, the data in Fig.~\ref{FormFactor}(b) shows no deviation from a purely exponential behavior which may arise from multiband superconductivity~\cite{Cubitt:2003ab} or Pauli paramagnetic effects effects~\cite{DeBeerSchmitt:2007aa,Bianchi:2008aa,White:2010aa}.

The rocking curve width ($w$) is inversely proportional to the longitudinal VL correlation length.
As shown in the Fig.~\ref{FormFactor}(b) inset, $w$ decreases with increasing field, approaching the experimental resolution.
This gradual ordering is commonly observed in superconductors with low pinning and attributed to an enhanced vortex-vortex interactions, and resulting increasing VL tilt modulus, as the density increases~\cite{Brandt:1995aa}.
This provides additional support for weak vortex pinning previous reported for {\CsTa}~\cite{Li.2022cod}.

\subsection*{Temperature dependence of scattering intensity}
\begin{figure}
    \includegraphics{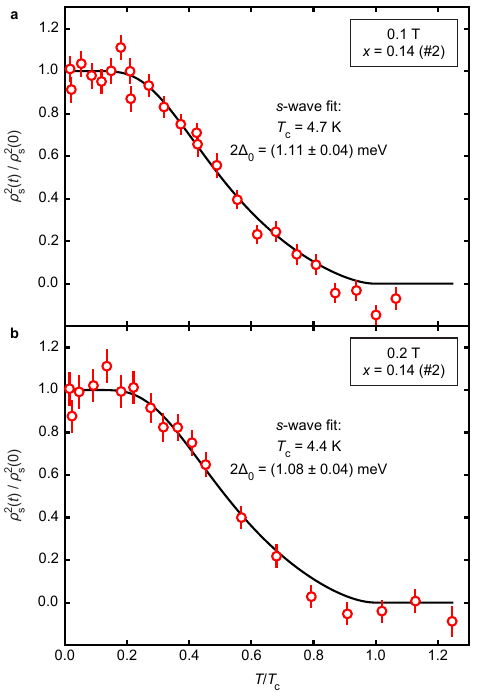}
    \caption{\label{TempDep}
        {\bf Temperature dependence of the superfluid density.}
        {\bf a} $0.1$~T and {\bf b} $0.2$~T.
        Error bars indicate one standard deviation.
        The curves are fits to an $s$-wave model as described in the text, with gap values and critical temperatures indicated in the plots.
        }
\end{figure}

The form factor is proportional to the superfluid density ($\rho_s$) which is dominated by the lowest gap values for a reduced temperature $t = T/T_c \lesssim \tfrac{1}{3}$.
Figure~\ref{TempDep} shows the normalized superfluid density versus temperature for two different applied fields, which display a clear saturation as $t \rightarrow 0$.
In the simplest case of an $s$-wave superconductor with a uniform gap
\begin{equation}
  \rho_s(t) = 1 - \frac{1}{2t} \int_0^{\infty} \cosh^{-2} \left( \frac{\sqrt{\varepsilon^2 + \Delta^2(t)}}{2t} \right) d \varepsilon,
  \label{rhos}
\end{equation}
where $\Delta(t)$ is the temperature dependent superconducting gap in units of $k_B T_c$~\cite{Prozorov:2006cd}.
In the weak coupling limit
\begin{equation}
  \Delta(t) = \Delta_0 \tanh \left( \frac{\pi}{\Delta_0} \sqrt{\frac{1}{t} - 1} \right),
  \label{GapTDep}
\end{equation}
and $\Delta_0$ is the zero-temperature gap amplitude~\cite{Gross:1986aa}.
Curves in Fig.~\ref{TempDep} show fits to the SANS data across the entire measured temperature range, using eqs.~(\ref{rhos}) and (\ref{GapTDep}) and the measured values for $T_c$.
This yields a superconducting gap of $2\Delta_0 = (1.11 \pm 0.04)~\mbox{meV}$ ($0.1$~T) and $(1.08 \pm 0.04)~\mbox{meV}$ ($0.2$~T) or $2\Delta_0/k_B T_c = 2.74 \pm 0.09$ and $2.85 \pm 0.11$ respectively, somewhat lower than the BCS prediction of $3.53$ and confirming weak coupling superconductivity.
This is consistent with the uniform gap across all FS sheets obtained from ARPES~\cite{Zhong.2023},
although our values are roughly 25\% smaller.
Estimates of the gap obtained from $\rho_s(t)$ determined by measurements of the lower critical field for $T \geq \tfrac{1}{3} T_c$ yielded an even larger gap, $2\Delta_0 = (3.6 \pm 0.8)~\mbox{meV}$~\cite{Li.2022cod}.
We note that a uniform (nodeless) gap does not preclude e.g. $s \pm is$ pairing that would break time-reversal symmetry or multiband superconductivity with a $s^{++}$ or $s^{+-}$ state~\cite{Ritz.2023}, but is inconsistent with $d$-wave or $p$-wave pairing.

In summary, our SANS studies of the {\CsTa} VL indicates a wholly conventional superconducting state.
Furthermore, the optimal superconductivity without CDW order is likely a conventional BCS superconductor, where electron pairing is induced by electron-lattice coupling.
Our results on {\CsTa} thus suggest that the  exotic properties reported for CDW related phenomena in other members of the $A$V$_3$Sb$_5$ superconductors are likely not related to the microscopic origin of superconductivity.

\section*{Methods}
Single crystals of Cs(V$_{1-x}$Ta$_x$)$_3$Sb$_5$ were grown by by the self-flux method~\cite{Ortiz:2019dl,Ortiz:2020by,Wang:2021gh,Zhong.2023}. 
These materials form in a layered kagome structure with a $P6/mmm$ space group~\cite{Ortiz:2020by}, and with facets that allow for an easy determination of the in-plane crystalline axes.
For the SANS measurements a mosaic of co-aligned {\CsTa} single crystals with a critical temperature $\Tc = 5.3$~K and a total mass of 200~mg was used, oriented with the $[1 \, 0 \, 0]$-axis horizontal and the $[\overline{1} \, 2 \, 0]$-axis vertical.
The co-alignment of the individual crystals in the mosaics is confirmed by the six sharp VL Bragg peaks observed in Figs.~\ref{SampleDifPats}(c)-\ref{SampleDifPats}(f) and the absence of significant scattered intensity in between those.
This also excludes twinning within the single crystals.

The SANS measurements were carried out using the SANS-I instrument at the Swiss Spallation Neutron Source (SINQ) at the Paul Scherrer Institute.
Two experiments were performed, using a pumped $^4$He cryomagnet for measurements down to $1.7$~K~(\#1) or a dilution refrigerator (DR) for measurements between 67~mK and $T_c$~(\#2).
A reference measurement was performed on a mosaic of undoped {\Cs} sample (280~mg, $\Tc = 3.0$~K) using the GP-SANS instrument at the High Flux Isotope Reactor (HFIR) at Oak Ridge National Laboratory (ORNL).
The measurements were carried out at a temperature of 35~mK using a DR.

In all cases, a neutron wavelength $\lambda_n = 1.4$~nm and bandwidth of $\Delta \lambda_n/\lambda_n = 10\%$ was used, and the diffracted neutrons were detected using a position-sensitive detector placed at 11~m from the sample.
The horizontal magnetic field was applied along the crystalline $[0 \, 0 \, 1]$-direction and near-parallel to the incident neutron beam.
The sample and cryomagnet were rotated together about the horizontal axes perpendicular to the beam direction to satisfy the Bragg condition for the different VL peaks.
Small-angle background measurements were collected in zero field at the base temperature for the respective experiment, and subtracted from the data.

All SANS data was analyzed using the GRASP graphical reduction and analysis software for small-angle neutron scattering available at https://www.ill.fr/grasp/~\cite{Dewhurst.2023}.

\section{Acknowledgments}
The neutron scattering work was supported by the U.S. Department of Energy, Office of Basic Energy Sciences, under awards no. DE-SC0005051 (N.C., M.R.E.) and no. DE-SC0012311 (Y.X, W.Y, P.D), and by the Swiss National Science Foundation (SNSF) Project grant 200021\_188707 (J.S.W.).
The single-crystal characterization efforts at Rice are supported by the Robert A. Welch Foundation Grant No. C-1839 (P.D.).
The work at Beijing Institute of Technology (BIT) was supported by the National Key R\&D Program of China Grants No. 2020YFA0308800 and 2022YFA1403400, and by the Beijing Natural Science Foundation Grant No. Z210006 (Z.W., J.L, Y.Y.).
Z.W. thanks the Analysis \& Testing Center at BIT for assistance in facility support. 
This work is based on experiments performed at the Swiss Spallation Neutron Source SINQ, Paul Scherrer Institute, Villigen, Switzerland.
A portion of this research used resources at the High Flux Isotope Reactor, a DOE Office of Science User Facility operated by the Oak Ridge National Laboratory.

\section{Author contributions}
M.R.E., P.D. and J.-X.Y. conceived and designed the experiments.
Y.X., N.C. W.Y., L.M.D.-S, J.S.W., P.D., and M.R.E. performed the experiments.
Y.X. and N.C. analyzed the data.
Z.W., J.L., and Y.Y. contributed materials/analysis tools.
M.R.E., Y.X, N.C., and P.D. wrote the paper.

\bibliography{Cs135}

\end{document}